\documentclass[prb,showpacs,preprintnumbers,amsfonts,amssymb,amsmath,floats,twocolumn,aps]{revtex4}

\usepackage{graphicx}
\usepackage{dcolumn}
\usepackage{bm}
\usepackage[final,dvips]{epsfig}
\usepackage{bm}
\usepackage{color}

\newcommand{\ff}[1]{{\bm #1}}
\newcommand{\ca}[1]{{\cal #1}}
\newcommand{\tr}{\mbox{tr}}

\newcommand{\up}{{\scriptscriptstyle\wedge}}
\newcommand{\lo}{{\scriptscriptstyle\vee}}
\newcommand{\ocite}[1]{Ref.\ \onlinecite{#1}}
\newcommand{\refeq}[1]{Eq.\ (\ref{eq:#1})}
\newcommand{\labeq}[1]{\label{eq:#1}}
\newcommand{\reffig}[1]{Fig.\ \ref{fig:#1}}
\newcommand{\labfig}[1]{\label{fig:#1}}

\begin{document} 
  
\title{Non-equilibrium cluster-perturbation theory} 

\author{Matthias Balzer and Michael Potthoff}

\affiliation{
I. Institut f\"ur Theoretische Physik,
Universit\"at Hamburg, 
Germany
}
 
\begin{abstract}
The cluster perturbation theory (CPT) is one of the simplest but systematic quantum cluster approaches to lattice models of strongly correlated electrons with local interactions. 
By treating the inter-cluster potential, in addition to the interactions, as a perturbation, it is shown that the CPT can be reformulated as an all-order re-summation of diagrams within standard weak-coupling perturbation theory where vertex corrections are neglected.
This reformulation is shown to allow for a straightforward generalization of the CPT to the general non-equilibrium case using contour-ordered Green's functions.
Solving the resulting generalized CPT equation on the discretized Keldysh-Matsubara time contour, the transient dynamics of an essentially arbitrary initial pure or mixed state can be traced.
In this way, the time-dependent expectation values of one-particle observables can be obtained within an approximation that neglects spatial correlations beyond the extension of the reference cluster.
The necessary computational effort is very moderate. 
A detailed discussion and simple test calculations are presented to demonstrate the strengths and the  shortcomings of the proposed approach.
The non-equilibrium CPT is systematic and is controlled in principle by the inverse cluster size. 
It interpolates between the non-interacting and the atomic or decoupled-cluster limit which are recovered exactly and is found to predict the correct dynamics at very short times in a general non-trivial case. 
The effects of initial-state correlations on the subsequent dynamics and the necessity to extend the Keldysh contour by the imaginary Matsubara branch are analyzed carefully and demonstrated numerically. 
It is furthermore shown that the approach can describe the dissipation of spin and charge to an uncorrelated bath with an essentially arbitrary number of degrees of freedom. 
\end{abstract} 
 


\maketitle 

\section{Introduction}

A theoretical understanding of transient processes in systems of strongly correlated electrons far away from thermal equilibrium and the development of according methods is one of the most challenging tasks in condensed-matter physics.
There is a pure theoretical motivation, on the one hand, since the study of non-equilibrium states opens up a new perspective on classical many-body effects, such as collective magnetic order, high-temperature superconductivity, Kondo screening of local moments or Mott metal-insulator transitions, for example.
On the other hand, there is an urgent need to describe and understand the results of recent exciting experimental studies in different fields: 
This includes nanostructure physics as, for example, the application of scanning-tunnelling microscope techniques to measure the spin relaxation time of itinerant and correlated electrons in nanostructures with atomic resolution, \cite{LEL+10}
or relaxation and switching times in first atom-by-atom realizations of all-spin based spintronics devices. \cite{KWCW11}
Furthermore, an improved theoretical understanding of fast demagnetization processes probed by femtosecond optical excitations \cite{MWD+09} and of the non-equilibrium electronic structure of strongly correlated transition-metal oxides which is accessible to femtosecond pump-probe spectroscopies. \cite{PLL+06,WPBC09}
Another fascinating field is the controlled preparation and monitoring of the non-equilibrium dynamics of highly excited fermionic states realized in correlated systems of ultracold atoms in optical lattices. \cite{SGJ+10}
In all these examples, the most interesting questions refer to the effect of strong nonlocal electronic correlations on the dynamics of itinerant electrons on a lattice or in a well-defined nanostructure.

For the strong-correlation regime of extended systems, non-perturbative numerical methods are required. 
Besides exact-diagonalization techniques \cite{PL86} which are limited to systems with small Hilbert-space dimensions, numerical renormalization-group \cite{AS05} or density-matrix renormalization-group techniques \cite{WF04} can be used to study impurity or one-dimensional lattice systems with high numerical accuracy.
The continuous-time quantum Monte-Carlo approach can straightforwardly be extended to the non-equilibrium case. \cite{EKW09}
It also belongs to the class of numerically exact methods but is limited, due to the sign or phase problem, to short-time dynamics.
Among the non-perturbative but approximate techniques, Green's function-based embedding methods are attractive. 
Relying on the pioneering work of Kubo, \cite{Kub57} Schwinger, \cite{Sch61} Kadanoff, Baym \cite{BK61} and Keldysh, \cite{Kel65}
(see also Refs.\ \onlinecite{Dan84,RS86,Wag91}) all-order diagrammatic re-summations can be used to define non-equilibrium generalizations of dynamical mean-field theory, \cite{FTZ06,SM02} self-energy functional theory \cite{PB11} or of the dual-fermion approach. \cite{JLB+10}
All of the above-mentioned impurity or cluster-embedding methods are highly expensive numerically. 

The purpose of the present paper is to propose and to discuss a method which is obtained by a generalization of the cluster-perturbation theory (CPT). \cite{SPPL00,SPP02,GV93,ZEAH02}
This non-equilibrium CPT is a conceptually simple method which can be applied to lattice models of correlated electrons with local interactions and basically arbitrary initial states and arbitrary Hamiltonian dynamics. 
The required computational resources are very moderate.
It is based on a partitioning of the lattice model of interest into smaller parts (``clusters'') that are amenable to an exact solution, preferably by means of exact diagonalization, and treats the initially disregarded inter-cluster terms subsequently in an approximative way such that the method becomes systematic and controlled by the inverse cluster size.
The non-equilibrium CPT accounts for temporal correlations and includes non-local but short-range spatial correlations up to the scale of the cluster size in the spirit of cluster mean-field methods. \cite{MJPH05}
It is thereby closely related to the (cellular) dynamical mean-field approach, and can be seen as the starting point for more elaborate but also more expensive self-energy-functional or dual-fermion techniques.
The proposed non-equilibrium CPT is the simplest systematic approach to non-equilibrium dynamics which includes non-local correlations.

Our formal idea is to first re-construct the usual equilibrium CPT by means of the standard weak-coupling perturbation expansion but treating besides the bilinear inter-cluster hopping the quartic interaction terms as a perturbation as well.
The CPT Green's function is then obtained by formally summing all diagrams to infinite order but neglecting certain vertex corrections.
In a second step, this idea can straightforwardly be transferred to the non-equilibrium situation by replacing the thermal Green's function with the contour-ordered Green's function.
The central CPT equation thereby becomes a matrix equation in orbital and (discretized) time indices which can easily be solved numerically. 
 
The paper is organized as follows:
The basic theory of non-equilibrium Green's functions is reviewed in the next section \ref{sec:ne} with notations following \ocite{Wag91}. 
Section \ref{sec:cpt} develops the non-equilibrium cluster-perturbation theory in detail. 
An extensive discussion of the new approach and of different numerical results is given in section \ref{sec:dis}.
The conclusions are summarized in section \ref{sec:con}.

\section{Expansion of the non-equilibrium Green's function}
\label{sec:ne}

Consider a system of electrons which at time $t_0$ is in a normalized pure state $|{\Psi}\rangle$.
We assume that this state is the $N$-particle ground state of some properly defined Hamiltonian 
\begin{equation}
  B = B_0 + B_1 \: ,
\end{equation}
where $B_0$ is a one-particle operator and $B_1$ an interaction term. 
Alternatively, the system could be at time $t_0$ in a mixed state $\rho$ where it is assumed that a Hamiltonian $B$ can be found such that
\begin{equation}
  \rho = \frac{\exp(-\beta \ca B) }{ \tr \exp(-\beta \ca B)} \: ,
\end{equation}
where $\ca B = \ca B_0 + B_1 = B -\mu N$ and where $\beta$ is the inverse temperature of the initial state.
With $\rho=|{\Psi}\rangle \langle{\Psi}|$ and $\beta \to \infty$ this also comprises pure initial states.

For $t>t_0$ the system's time evolution shall be governed by the explicitly time-dependent Hamiltonian
\begin{equation}
  H(t) = H_0 + H_1(t) \: ,
\end{equation}
where $[H(t),B]_-\ne 0$ in general.
For the calculations below, we will assume that the system is not driven by explicitly time-dependent external fields and that $H(t) = H(0) \ne B$. 
However, the formalism will be developed for the general case.

Consider an arbitrary possibly time-dependent observable $A(t)$.
Its time dependence within the Heisenberg picture with respect to 
$\ca H(t) = H(t) - \mu N$ is determined by the equation of motion
\begin{equation}
  i \frac{d}{dt} A_{\ca H}(t) = [A_{\ca H}(t),\ca H(t)]_- + i \frac{\partial}{\partial t} A_{\ca H}(t) 
\end{equation}
with the initial condition $A_{\ca H}(t_0) = A(t_0)$.
The formal solution of the equation of motion is given by 
\begin{equation}
A_{\ca H}(t) 
= 
\left[ 
\overline{\ca T} e^{
i \int_{t_0}^{t} d\widetilde{t} \ca H(\widetilde{t}) 
}
\right] 
A(t)
\left[ \ca T e^{
- i \int_{t_0}^t d\widetilde{t} \ca H(\widetilde{t}) 
}
\right] 
\: ,
\end{equation}
where $\ca T$ ($\overline{\ca T}$) is the chronological (anti-chronological) time-ordering operator.

\begin{figure}[t!]
\centerline{\includegraphics[width=0.85\columnwidth]{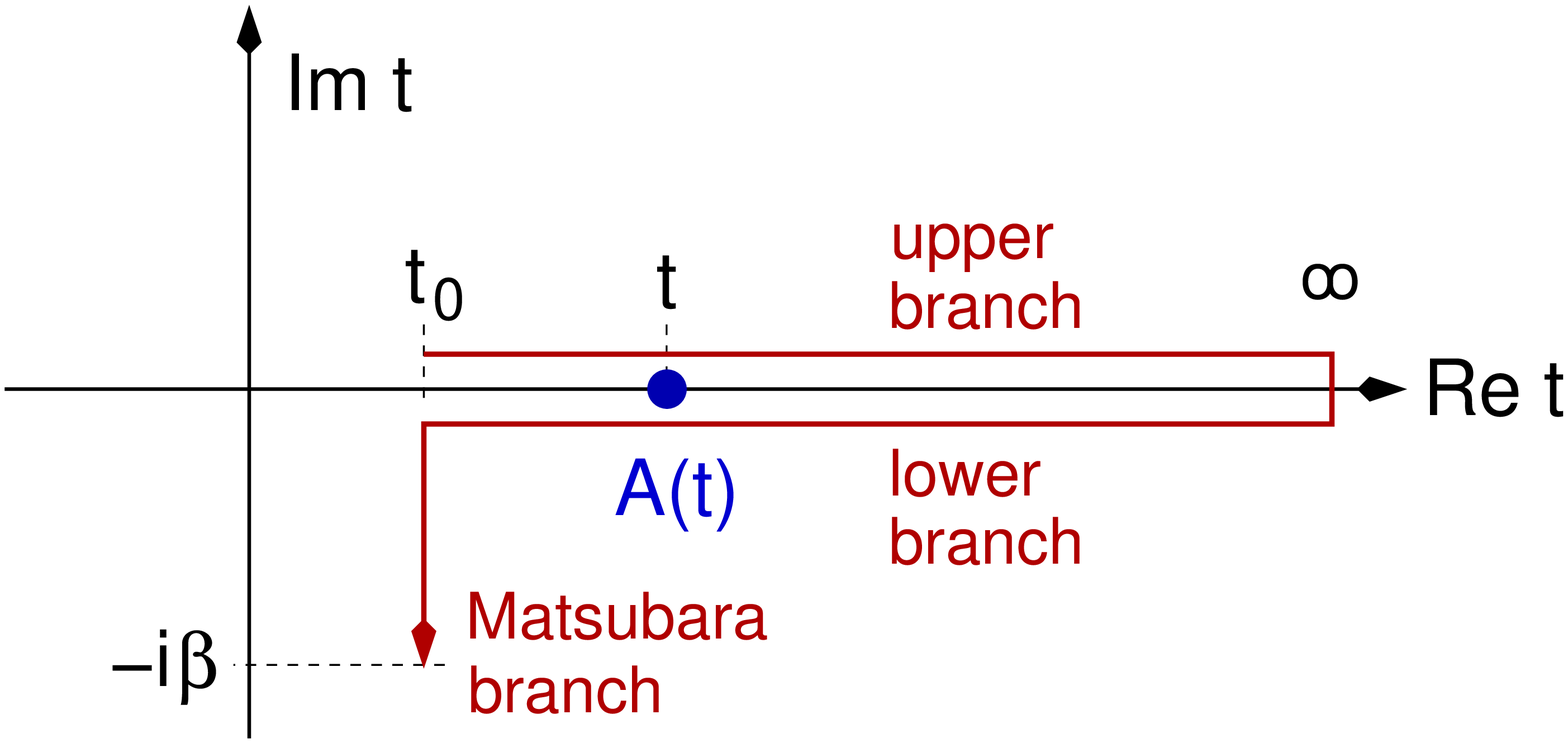}}
\caption{ 
Three-branch contour $\gamma$ in the complex time plane. The upper and the lower real branches define the Keldysh contour, the imaginary branch is called the Matsubara branch.
}
\labfig{cont}
\end{figure}

For a system in the initial state $\rho$ the expectation value of the observable $A(t)$ at time $t$ is 
$\langle A \rangle_t = \tr (\rho A_{\ca H}(t))$.
This can be written as: \cite{Wag91}
\begin{equation}
\langle A \rangle_t 
=
\frac{
\tr \left(
\ca T_\gamma \exp\left(
- i \int_\gamma d\widetilde{t} \, \ca K(\widetilde{t}) 
\right) A(t) \right)
}{
\tr \left(
\ca T_\gamma \exp\left(
- i \int_\gamma d\widetilde{t} \, \ca K(\widetilde{t}) 
\right) \right)
} \: .
\end{equation}
Here, the time integration is carried out along the contour $\gamma$ in the complex time plane. 
$\gamma$ extends from $\widetilde{t}=t_0$ to $\widetilde{t}=\infty$ along the real axis (upper branch) and back from $\widetilde{t}=\infty$ to $\widetilde{t}=t_0$ along the real axis (lower branch) and finally from $\widetilde{t}=t_0$ to $\widetilde{t}=t_0-i \beta$ along the imaginary axis (Matsubara branch), see \reffig{cont}.
We also refer to the upper and the lower branch as the Keldysh contour. 
$\ca T_\gamma$ denotes the ordering operator along the contour and, after expanding the exponential, places an operator $\ca K(t_1)$ to the left of $\ca K(t_2)$ if $t_1$ is ``later'' than $t_2$ on the contour $\gamma$ where $t_0 - i\beta$ is the ``latest'' time.
Obviously, $\ca T_\gamma$ replaces $\ca T$ on the upper and $\overline{\ca T}$ on the lower branch.
Finally, $\ca K(\widetilde{t}) = \ca H(\widetilde{t})$ on the upper and the lower branch of $\gamma$ while $\ca K(\widetilde{t}) = \ca B$ on the Matsubara branch.

$\ca T_\gamma$ also acts on $A(t)$.
The time argument of $A(t)$ is the time at which the expectation value is evaluated and indicates the position on the time contour where for the integrals in the numerator the observable has to be placed.
Note that, for the numerator, the results of integrating along the upper and the lower branches between $\widetilde{t}=t$ and $\widetilde{t}=\infty$ cancel each other. 
Hence, the integration along the Keldysh part of the contour can be limited to $\widetilde{t} < t$.
For the denominator, only the Matsubara branch of the contour contributes to the integral with the result $\tr \exp(-\beta \ca B)$.
If $H$ is time-independent and equal to $B$, the equilibrium result $\langle A \rangle_t = \tr (\exp(-\beta H) A(t)) / \tr \exp(-\beta H)$ is recovered.

We assume the Hamiltonian $B$, which characterizes the initial state, and the Hamiltonian $H(t)$, which determines the system's dynamics, to be given in second-quantized form:
\begin{equation}
B  
= 
\sum_{\alpha\beta} T^{\rm (B)}_{\alpha\beta} c_{\alpha}^\dagger c_\beta
+ 
\frac{1}{2} \sum_{\alpha\beta\gamma\delta} U^{\rm (B)}_{\alpha\beta\delta\gamma} 
c_{\alpha}^\dagger c_\beta^\dagger c_\gamma c_\delta
\labeq{b2q}
\end{equation}
and 
\begin{equation}
H(t) 
= 
\sum_{\alpha\beta} T_{\alpha\beta} c_{\alpha}^\dagger c_\beta
+ 
\frac{1}{2} \sum_{\alpha\beta\gamma\delta} U_{\alpha\beta\delta\gamma}(t) c_{\alpha}^\dagger c_\beta^\dagger c_\gamma c_\delta \: .
\labeq{h2q}
\end{equation}
Here $\alpha$ refers to a complete and orthonormal set of (time-independent) one-particle orbitals, i.e.\ the explicit time-dependence is due to the interaction parameters only.
An external bilinear time-dependent field could be considered in addition. 
In this case the interaction part would also contain terms bilinear in $c_\alpha$ and $c_\alpha^\dagger$.

The time-dependent expectation value of any one-particle observable $A(t)=\sum_{\alpha\beta} a_{\alpha\beta}(t) c_\alpha^\dagger c_\beta$ can be obtained from the contour-ordered Green's function
\begin{equation}
 i G_{\alpha\alpha'}(t,t') = \langle \ca T_\gamma c_{\ca K,\alpha}(t) c^\dagger_{\ca K,\alpha'}(t') \rangle
\end{equation}
as 
\begin{equation}
  \langle A \rangle_t = -i \sum_{\alpha\beta} a_{\alpha\beta}(t) G_{\beta\alpha}(t,t+0^+)
\labeq{obs}
\end{equation} 
where $0^+$ is a positive infinitesimal and $\langle \cdots \rangle = \tr (\rho \cdots)$ denotes the expectation value in the initial state.
Furthermore, the annihilator and the creator are given in the Heisenberg picture with respect to $\ca K(t)$, $t,t'$ are arbitrary times on the contour, and $\ca T_\gamma$ is the time ordering of annihilators and creators on the contour $\gamma$ which yields an additional (Fermi) sign per transposition.

The contour-ordered Green's function involves operators given in the Heisenberg picture, i.e.\ with a  time-dependence due to the {\em interacting} Hamiltonian $\ca H$, and an expectation value with a (mixed) state corresponding to the {\em interacting} Hamiltonian $\ca B$.
The main motivation for placing the contour-ordered Green's function in the focus of the theory, rather than, for example, expectation values like $\langle A \rangle_t$, is that (i) the Green's function can be brought into a form that meets the requirements to apply Wick's theorem and that (ii) the application of Wick's theorem only generates contour-ordered Green's functions again. 
Thereby, a closed set of physically interesting quantities is obtained, and a consistent perturbation theory can be set up.

Following \ocite{Wag91}, the contour-ordered Green's function can be cast into the form:
\begin{equation}
i G_{\alpha\alpha'}(t,t') 
=
\frac{
\langle
\ca T_\gamma \: 
e^{
- i \int_\gamma d\widetilde{t} \, \ca K_{\ca K_0,1}(\widetilde{t}) 
} 
c_{\ca K_0,\alpha}(t) c^\dagger_{\ca K_0,\alpha'}(t') 
\rangle^{(0)}
}{
\langle
\ca T_\gamma \:
e^{
- i \int_\gamma d\widetilde{t} \, \ca K_{\ca K_0,1}(\widetilde{t}) 
} 
\rangle^{(0)}
} 
\: .
\end{equation}
In this expression, the annihilators and creators, $c_{\ca K_0,\alpha}(t)$ and $c^\dagger_{\ca K_0,\alpha'}(t')$ possess a ``free'' time dependence only, i.e.\ they are given in the interaction picture where the time dependence is due to $\ca K_0$ only. 
The same applies to the interaction term $\ca K_{\ca K_0,1}(\widetilde{t})$ appearing under the contour integral -- its time dependence is ``free'' and given by $\ca K_0$ only. 
Finally, also the expectation value $\langle \cdots \rangle^{(0)} = \tr (\rho_0 \cdots)$ is a ``free'' one and is defined with free density operator $\rho_0 = \exp(-\beta \ca B_0) / Z_0$ only. 
Hence, we can apply Wick's theorem and therewith standard techniques of perturbation theory.

Expanding the Green's function in powers of the interaction parts of $\ca B$ and $\ca H$, the $n$-th order coefficient turns out to be given in terms of $2n+1$ ``free'' contour-ordered Green's functions:
\begin{equation}
 i G^{(0)}_{\alpha\alpha'}(t,t') 
 = 
 \langle \ca T_\gamma c_{\ca K_0,\alpha}(t) c^\dagger_{\ca K_0,\alpha'}(t') \rangle^{(0)} \: .
\end{equation}
This can be computed exactly for the case considered here, i.e.\ for $\ca H_0(t) = \ca H_0 = \mbox{const.}$ but $[B_0,H_0]_-\ne 0$.
We find:
\begin{equation}
i G^{(0)}_{\alpha\alpha'}(t,t') 
=
\left( 
e^{-i (\ff T_{\rm K} -\mu) t} 
\frac{1}{1+e^{-\beta(\ff T_{\rm B} - \mu)}} 
e^{i(\ff T_{\rm K} -\mu) t'} 
\right)_{\alpha{\alpha'}}
\labeq{gfree1}
\end{equation}
if $t$ later than $t'$ on $\gamma$ and
\begin{equation}
i G^{(0)}_{\alpha\alpha'}(t,t') 
= -
\left( 
e^{-i(\ff T_{\rm K} -\mu) t} 
\frac{1}{e^{\beta(\ff T_{\rm B} - \mu)}+1}
e^{i(\ff T_{\rm K} -\mu) t'} 
\right)_{\alpha{\alpha'}}
\labeq{gfree2}
\end{equation}
if $t'$ later than $t$ on $\gamma$.
On the Keldysh contour $t$ is real and $\ff T_{\rm K}=\ff T$ with the elements $T_{\alpha\beta}$ while on the Matsubara branch $t=-i\tau$ with $0\le \tau \le \beta$ and $\ff T_{\rm K}=\ff T_{\rm B}$ with elements $T^{\rm (B)}_{\alpha\beta}$, see \refeq{b2q} and \refeq{h2q}.

\section{Cluster-perturbation theory}
\label{sec:cpt}

There are several ways to define the cluster-perturbation theory (CPT) for the equilibrium case. 
The first approach, based on the so-called Hubbard-I approximation, \cite{Hub63} focuses on the electron self-energy of the Hubbard model \cite{Hub63,Gut63,Kan63} for a $D$ dimensional lattice.
The Hubbard-I approximation can be constructed by starting from the atomic limit of the Hubbard model and taking the self-energy from that limit as an approximation for the infinite lattice model.
In the original work, \cite{Hub63} additional requirements on the average occupation numbers are imposed which must be solved self-consistently.
The Hubbard-I concept was generalized later \cite{GV93} by starting from a finite Hubbard cluster instead of a single Hubbard atom.
Approximating the lattice self-energy by the self-energy of a cluster consisting of a finite number of $L_c$ sites, defines a numerical technique which (i) directly works in the thermodynamical limit, (ii) can be improved systematically by increasing the cluster size $L_c$ and (iii) provides, via Dyson's equation, a single-electron Green's function which respects certain general requirements of Lehmann representability and causality.
On the other hand, this construction of the CPT appears to be rather {\em ad hoc}.

The second approach is based on strong-coupling perturbation theory for the Hubbard model and is more systematic. 
For Hubbard-type models, an expansion in powers of the hopping $t$ around the atomic limit can be organized in a systematic diagrammatic series. \cite{Met91,PST98,RKL08} 
At the lowest order this leads to the Hubbard-I approximation. 
The CPT is obtained from a cluster generalization of the strong-coupling expansion.
The extension consists in a partitioning of the lattice into small clusters that can be treated exactly, and a subsequent expansion in powers of the inter-cluster hopping.
The lowest order constitutes the CPT. \cite{SPPL00,SPP02}
In principle, the expansion can be carried out to arbitrary order in the inter-cluster hopping using the diagrammatic method of Refs.~\onlinecite{Met91,PST00} or the cluster dual-fermion method. \cite{HBR+07}
However, going beyond the lowest order is quite demanding numerically and leads to causality problems at large $t$ and low temperatures due to the degeneracy of the ground state.
Since the lowest order of the strong-coupling expansion is causal and still represents a systematic approach with respect to the cluster size $L_c$, it has gained some attraction in the past. 
The CPT is a conceptually simple method which nevertheless includes short-range correlations on the scale of the cluster size and which requires moderate computational resources only.

An alternative approach to construct the CPT is presented here.
It is based on the usual weak-coupling perturbation expansion.
Besides the quartic interaction terms in $\ca B$ and $\ca H(t)$, however, we additionally treat the bilinear inter-cluster hopping as a perturbation as well.
The CPT Green's function is then obtained by formally summing all diagrams to infinite order but neglecting vertex corrections.
This idea can straightforwardly be transferred to the non-equilibrium situation by replacing the thermal Green's function with the contour-ordered Green's function.

\begin{figure}[tb]
\centerline{\includegraphics[width=0.5\columnwidth]{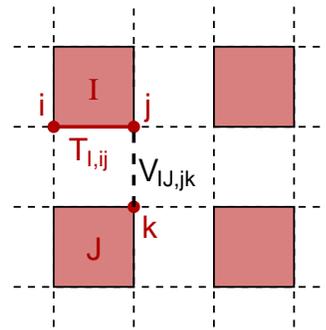}}
\caption{ 
Partitioning of a $D=2$ square lattice into clusters with $L_c=4$ sites each. 
$T_{I,ij}$ denotes the intra-cluster hopping between sites $i$ and $j$ within the same cluster $I$.
$V_{IJ,ij}$ is the inter-cluster hopping between sites $i\in I$ and $j\in J$.
}
\labfig{cpt}
\end{figure}

Starting point for the construction of the CPT is a partitioning of the original $D$ dimensional lattice consisting of $L$ sites into clusters of finite size and open boundaries.
The clusters shall consist of $L_c$ sites each. 
Fig.\ \reffig{cpt} gives an example for the $D=2$ square lattice and $L_c=4$.
For simplicity, we assume all clusters to be identical and to form a superlattice labeled by a superlattice site index $I=1,...,L/L_c$. 
The sites within the cluster $I$ are labeled by an index $i=1, ..., L_c$. 

The Hamiltonians of the initial thermal and of the transient final state, i.e.\ $B$ and $H(t)$, are decomposed accordingly,
\begin{equation} 
  B = B' + B_{\rm inter} 
  \; , \qquad
  H(t) = H'(t) + H_{\rm inter} \: .
\end{equation} 
$B'$ and $H'(t)$ correspond to the reference system of decoupled clusters. 
We have
\begin{equation} 
  B' = \sum_{I=1}^{L/L_c} B'_I
  \; , \qquad
  H'(t) = \sum_{I=1}^{L/L_c} H'_I(t) \: ,
\end{equation} 
where $B'_I$ and $H'_I(t)$ describe the thermal initial state and the dynamics of the isolated cluster $I$.
The CPT is mainly designed for applications to Hubbard-type models with local interactions.
Besides the intra-cluster hopping, we therefore assume the interaction terms $B_1$ and $H_1(t)$ to be fully included in the reference system.
Hence:
\begin{equation} 
  B'_I
  = 
  \sum_{i,j=1}^{L_c} \sum_{\sigma_i \sigma_j} 
  \varepsilon^{\rm (B)}_{I,ij\sigma_i\sigma_j} 
  c_{Ii\sigma_i}^\dagger c_{Ij\sigma_j}
  +
  B_{1,I}
\end{equation} 
and
\begin{equation} 
  H'_{I}(t)
  = 
  \sum_{i,j=1}^{L_c} \sum_{\sigma_i \sigma_j} 
  \varepsilon_{I,ij\sigma_i\sigma_j}
  c_{Ii\sigma_i}^\dagger c_{Ij\sigma_j}
  +
  H_{1,I}(t)
  \: ,
\end{equation} 
where $i$ and $j$ run over the sites within the cluster $I$, where $\sigma_i$ labels the residual orbital and spin degrees of freedom at a site $i$, and where $B_{1,I}$ and $H_{1,I}(t)$ denote the respective interaction part within cluster $I$.
On the other hand, the inter-cluster parts include bilinear hopping terms only:
\begin{equation} 
  B_{\rm inter}
  = 
  \sum_{I,J}^{I \ne J} 
  B_{\rm inter, IJ}
  \; ,
  \qquad
  H_{\rm inter}
  = 
  \sum_{I,J}^{I \ne J} 
  H_{\rm inter, IJ}
\end{equation} 
where
\begin{equation} 
  B_{\rm inter, IJ}
  = 
  \sum_{i\in I,j\in J} \sum_{\sigma_i \sigma_j} 
  V^{\rm (B)}_{IJ,ij,\sigma_i\sigma_j}
  c_{i\sigma_i}^\dagger c_{j\sigma_j}
  \: .
\end{equation} 
and
\begin{equation} 
  H_{\rm inter, IJ}
  = 
  \sum_{i\in I,j\in J} \sum_{\sigma_i \sigma_j} 
  V_{IJ,ij,\sigma_i\sigma_j}
  c_{i\sigma_i}^\dagger c_{j\sigma_j}
  \: .
\end{equation} 

A triple of indices $(I,i,\sigma_i)=\alpha$ labels a certain orbital of the one-particle basis.
With respect to this basis, the intra-cluster and the inter-cluster hopping parameters form matrices $\ff \varepsilon_{\rm B}$, $\ff V_{\rm B}$ and  $\ff \varepsilon$, $\ff V$, respectively. 
We have $\ff T_{\rm B}=\ff \varepsilon_{\rm B}+\ff V_{\rm B}$ and $\ff T=\ff \varepsilon+\ff V$, see
\refeq{b2q} and \refeq{h2q}.
In case that the superlattice of clusters is invariant under translations, Fourier transformation block-diagonalizes $\ff V_{\rm B}$ and $\ff V$ simultaneously. 
Exploiting the fact that the intra-cluster hopping is already diagonal in and independent of the superlattice index $I$, we get matrices of the form: $\ff \varepsilon_{\rm B}$, $\ff V_{\rm B}(\ff k)$ and $\ff \varepsilon$, $\ff V(\ff k)$, respectively, where $\ff k$ has the physical meaning of a wave vector and where $\ff \varepsilon_{\rm B}$ and $\ff \varepsilon$ are wave-vector independent. 
In all other cases, diagonalization must be done numerically, if desired. 
Note that $\ff T_{\rm B}$ and $\ff T$ are different for a general initial state and cannot be diagonalized simultaneously.

To set up the perturbation theory based on Wick's theorem, the quartic terms $B_1$ and $H_1(t)$ have to be treated as a perturbation.
As concerns the bilinear terms $\ca B_0$ and $\ca H_0$, however, we are free to treat them as ``free'' or as a ``perturbation''. 
Any choice is consistent with Wick's theorem.
A non-equilibrium generalization of the CPT is obtained when treating the inter-cluster couplings
$\ff V_{\rm B}$ and $\ff V$ as perturbations while $\ff \varepsilon_{\ff B}$ and $\ff \varepsilon$ are considered to be free.

Perturbation theory then provides us with Dyson's equation for the fully interacting contour-ordered Green's function:
\begin{equation} 
  \ff G = \ff G'_0 + \ff G'_0 \cdot \ff \Sigma_{\ff U_{\rm K}, \ff V_{\rm K}}[\ff G'_0] \cdot \ff G \: .
\labeq{dyson}
\end{equation} 
Here, all quantities are matrices with respect to time variables and orbital indices, such that the Green's function $\ff G$ has the elements $G_{\alpha\alpha'}(t,t')$, for example, and \refeq{dyson} is short for:
\begin{eqnarray} 
  G_{\alpha\alpha'}(t,t')
  &=& 
  {G_{\alpha\alpha'}^{(0)}}'(t,t') 
  + \sum_{\alpha''\alpha'''} \int_\gamma \int_\gamma dt'' dt'''
\nonumber \\
  &&{G_{\alpha\alpha''}^{(0)}}'(t,t'') 
  \Sigma_{\alpha''\alpha'''}(t'',t''') 
  G_{\alpha'''\alpha'}(t''',t') \; .
\nonumber \\
\labeq{dysonlong}
\end{eqnarray} 
The free Green's function $\ff G'_0$ in \refeq{dyson} is the $\ff U_{\rm B} = \ff U(t) = 0$, $\ff V_{\rm B} = \ff V = 0$ Green's function, i.e.\ the interaction-free intra-cluster contour-ordered Green's function or the interaction-free Green's function of the reference system. 
Explicitly, we have:
\begin{equation}
i {G^{(0)}_{\alpha\alpha'}}'(t,t') 
=
\left( 
e^{-i (\ff \varepsilon_{\rm K} -\mu) t} 
\frac{1}{1+e^{-\beta(\ff \varepsilon_{\rm B} - \mu)}} 
e^{i(\ff \varepsilon_{\rm K} -\mu) t'} 
\right)_{\alpha{\alpha'}}
\labeq{g01}
\end{equation}
if $t$ later than $t'$ on $\gamma$ and
\begin{equation}
i {G^{(0)}_{\alpha\alpha'}}'(t,t') 
= -
\left( 
e^{-i(\ff \varepsilon_{\rm K} -\mu) t} 
\frac{1}{e^{\beta(\ff \varepsilon_{\rm B} - \mu)}+1}
e^{i(\ff \varepsilon_{\rm K} -\mu) t'} 
\right)_{\alpha{\alpha'}}
\labeq{g02}
\end{equation}
if $t'$ later than $t$ on $\gamma$.
Here, $\ff \varepsilon_{\rm K}=\ff \varepsilon_{\rm B}$ if $t=-i\tau$ is on the Matsubara branch and $\ff \varepsilon_{\rm K}=\ff \varepsilon$ for real $t$ on the Keldysh contour.
The self-energy $\ff \Sigma_{\ff U_{\rm K}, \ff V_{\rm K}}[\ff G'_0]$ in \refeq{dyson} is obtained by summing over all irreducible self-energy insertions, formed by free propagators $\ff G'_0$ and vertices $\ff U_{\rm K}(t)$ and $\ff V_{\rm K}$ where $\ff U_{\rm K}(t) = \ff U_{\rm B}$ or $\ff U_{\rm K}(t) = \ff U(t)$, and likewise for $\ff V_{\rm K}$, depending on the position of the respective vertex on the time contour. 

The {\em exact} self-energy can formally be obtained in a two-step renormalization procedure, see \reffig{dyson}a. 
First, we consider the renormalization of the free propagators due to $\ff V_{\rm K}$, i.e.\ due to electron scattering at the non-local but instantaneous (local in time) inter-cluster potential. 
The corresponding self-energy is simply given by $\ff \Sigma_{\ff V_{\rm K}}[\ff G'_0] = \ff V_{\rm K} \otimes \ff 1$ with the $\delta$-function on the contour $1_{t,t'}=\delta_\gamma(t,t')$, and the renormalized propagator $\ff G_0$ is obtained as the solution of the corresponding Dyson equation:
\begin{equation} 
  \ff G_0 = \ff G'_0 + \ff G'_0 \cdot \ff V_{\rm K} \otimes \ff 1 \cdot \ff G_0  \: .
\labeq{dyson1}
\end{equation} 
This yields the Green's function for $\ff U_{\rm B} = \ff U(t) = 0$. 
Subsequent $\ff U_{\rm K}$ renormalization is formally achieved by introducing the corresponding self-energy $\ff \Sigma_{\ff U_{\rm K}}[\ff G_0]$ which is a (highly complicated) functional of the $\ff V_{\rm K}$-renormalized propagator.
This yields the full propagator as the solution of
\begin{equation} 
  \ff G = \ff G_0 + \ff G_0 \cdot \ff \Sigma_{\ff U_{\rm K}}[\ff G_0] \cdot \ff G \: .
\labeq{dyson2}
\end{equation} 
Since all diagrams are summed up, the procedure is exact. 
Comparison with Dyson's equation \refeq{dyson} shows that
\begin{equation} 
  \ff \Sigma_{\ff U_{\rm K}, \ff V_{\rm K}}[\ff G'_0]
  =  
  \ff \Sigma_{\ff U_{\rm K}}[({\ff G_0'}^{-1} - \ff V_{\rm K} \otimes \ff 1)^{-1}] 
  +
  \ff V_{\rm K}  \otimes \ff 1 \: .
\labeq{sigma}
\end{equation} 
Since the self-energy $\ff \Sigma_{\ff U_{\rm K}}[\ff G_0]$ is essentially unknown, this does not provide, of course, a pragmatic way to compute the full propagator.

\begin{figure}[tb]
\centerline{\includegraphics[width=0.8\columnwidth]{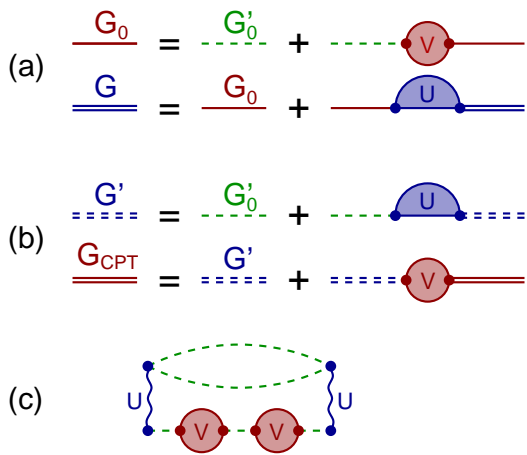}}
\caption{ 
Re-summation of diagrams generated by electron-electron scattering $U$ and by scattering at the inter-cluster potential $V$. 
(a) Exact procedure: Renormalization of the free ($U=V=0$) propagator $\ff G'_0$ by potential scattering followed by renormalization of the $U=0$ propagator $\ff G_0$ due to electron scattering [Eqs.\ (\ref{eq:dyson1}) and (\ref{eq:dyson2})].
(b) CPT: Renormalization of the free ($U=V=0$) propagator $\ff G'_0$ by electron scattering followed by renormalization of the $V=0$ propagator $\ff G'$ due to potential scattering [Eqs.\ (\ref{eq:dysoncpt1}) and (\ref{eq:dysoncpt2})].
(c) Self-energy diagram, second order in $U$, second order in $V$, which is not taken into account within CPT.
}
\labfig{dyson}
\end{figure}

Let us now consider the $\ff U_{\rm K}$ renormalization first, see \reffig{dyson}b.
This leads to the following Dyson equation:
\begin{equation} 
  \ff G' = \ff G'_0 + \ff G'_0 \cdot \ff \Sigma_{\ff U_{\rm K}}[\ff G'_0] \cdot \ff G' \: .
\labeq{dysoncpt1}
\end{equation} 
Its solution is the interacting Green's function $\ff G'$ of the reference system of decoupled clusters.
While still the functional form of $\ff \Sigma_{\ff U_{\rm K}}[\ff G'_0]$ is highly non-linear and unknown, the propagator $\ff G'$ may be calculated directly by exact diagonalization, provided that the cluster size $L_c$ is moderate.
Note that here it is essential to assume the $\ff U_{\rm K}$ vertex to be local and not to couple different clusters.
The subsequent $\ff V_{\rm K}$ renormalization of the already $\ff U_{\rm K}$-renormalized propagators is expressed with the Dyson equation
\begin{equation} 
  \ff G_{\rm CPT} = \ff G' + \ff G' \cdot \ff V_{\rm K} \otimes \ff 1 \cdot \ff G_{\rm CPT} \: .
\labeq{dysoncpt2}
\end{equation} 
Its solution defines the non-equilibrium CPT Green's function $\ff G_{\rm CPT}$. 
The reversed two-step renormalization is not exact since there are certain diagrams missing, see \reffig{dyson}c.
From \refeq{dysoncpt1} and \refeq{dysoncpt2} we get the CPT self-energy
\begin{equation} 
  \ff \Sigma_{{\rm CPT}, \ff U_{\rm K}, \ff V_{\rm K}}[\ff G'_0]
  =  
  \ff \Sigma_{\ff U_{\rm K}}[\ff G_0'] 
  +
  \ff V_{\rm K} \otimes \ff 1 \: .
\labeq{sigmacpt}
\end{equation} 
Comparing this expression with the exact self-energy \refeq{sigma} shows that CPT neglects the influence of scattering at the inter-cluster potential on the renormalization of propagators due to the interaction, i.e.\ vertex corrections. 
Another way to paraphrase the approximation is to say that the CPT neglects electron-electron ($\ff U_{\rm K}$) scattering {\rm across} different clusters but takes into account intra-cluster electron-electron scattering and scattering of electrons dressed by $\ff U_{\rm K}$ processes at the one-particle inter-cluster potential.

\section{Discussion and Results}
\label{sec:dis}

In the following we discuss the non-equilibrium CPT in detail and present numerical results to demonstrate, as a proof of principle, that the approach can be used in practice.

\subsection{Thermal equilibrium}

First, it has to be shown that the usual CPT is recovered for the case of thermodynamical equilibrium. 
We therefore assume that $H(t)=H=B$ for a moment.
Inspection of Eqs.\ (\ref{eq:g01}) and (\ref{eq:g02}) and of Eqs.\ (\ref{eq:gfree1}) and (\ref{eq:gfree2}) immediately shows that ${G_{\alpha\alpha'}^{(0)}}'(t,t')$ and ${G_{\alpha\alpha'}^{(0)}}(t,t')$ become temporally homogeneous, i.e.\ become functions of $t-t'$ only.
The interacting Green's function of the reference system, $G'_{\alpha\alpha'}(t,t')$, has to be computed exactly within non-equilibrium CPT and, therefore, is homogeneous.
Since $\ff V_{\rm K} \otimes \ff 1$ is homogeneous by definition, the CPT equation (\refeq{dysoncpt2}) proves the CPT Green's function $\ff G_{\rm CPT}$ to be homogeneous, too. 
With \refeq{obs} this implies that the expectation value of any (not explicitly time-dependent) observable $A$ is constant, $\langle A \rangle_t = \langle A \rangle_{t_0}$, and given by its thermal value for all $t>t_0$.

Furthermore, as is shown below, there is an independent CPT equation on the Matsubara branch only:
\begin{equation} 
  \underline{\ff G}_{\rm CPT} = \underline{\ff G}' + \underline{\ff G}' \cdot \ff V_{\rm B} \otimes \underline{\ff 1} \cdot \underline{\ff G}_{\rm CPT} \: .
\labeq{cptmat}
\end{equation} 
Here the underlined symbols represent matrices in $t,t'$ (besides orbital indices) {\em where $t,t'$ are restricted to the Matsubara branch only} and where the integrations implicit in the notations are limited accordingly. 
Together with the homogeneity of the quantities, this allows to transform to a Matsubara frequency representation:
\begin{equation} 
  {\ff G}_{\rm CPT}(i\omega_n) 
  = 
  {\ff G}'(i\omega_n) 
  + 
  {\ff G}'(i\omega_n) \ff V_{\rm B} {\ff G}_{\rm CPT}(i\omega_n) 
  \: ,
\end{equation} 
where $\omega_n =  (2n+1)\pi / \beta$ with integer $n$, and fat symbols stand for matrices with respect to orbital indices only.
After analytical continuation to arbitrary complex frequencies $i\omega_n \to \omega$, we therewith recover the usual equilibrium CPT equation \cite{SPPL00,SPP02} which may be solved by matrix inversion:
\begin{equation} 
  {\ff G}_{\rm CPT}(\omega) 
  = 
  \frac{\ff 1}{ {{\ff G}'(\omega) }^{-1} - \ff V_{\rm B}}
  \: ,
\end{equation} 
where translational symmetries of the lattice may be exploited by Fourier transformation in addition.

\refeq{cptmat} holds for the equilibrium but also for the general non-equilibrium case, i.e.\ for time inhomogeneous Green's functions.
Physically, it is a consequence of causality since the preparation of the initial state cannot depend on the subsequent time evolution of the system.

The CPT does respect this condition:
Consider an expression of the form
\begin{equation} 
  \ff I(t,t')
  =
  \int_\gamma dt'' \int_\gamma dt'''
  \ff G_1(t,t'') 
  \ff \Sigma(t'',t''') 
  \ff G_2(t''',t') \: ,
\labeq{tintegrals}
\end{equation} 
as it occurs in the Dyson equation (\ref{eq:dysonlong}) or, in a simpler form, in the CPT equation (\ref{eq:dysoncpt2}), and assume the external time variables $t$ and $t'$ to lie on the Matsubara branch. 
After integrating over $t'''$, the integrand for the remaining $t''$ integration depends on $t''$ and $t,t'$ only.
In particular, since $t,t'$ by assumption are always ``later'' than $t''$ on $\gamma$, if $t''$ is real, it does not matter whether $t''$ lies on the upper or on the lower branch of $\gamma$.
Therefore, the integration along the entire Keldysh branch does not contribute to the integral and
\begin{equation} 
  \ff I(t,t')
  =
  \int_{t_0}^{t_0-i\beta} dt'' \int_\gamma dt'''
  \ff G_1(t,t'') 
  \ff \Sigma(t'',t''') 
  \ff G_2(t''',t') \: .
\end{equation} 
Using the same arguments, we can then also replace
\begin{equation} 
  \int_\gamma dt'''
  \mapsto
  \int_{t_0}^{t_0-i\beta} dt'''
\end{equation} 
and we are left with integrations along the Matsubara branch only. 

\subsection{Time discretization}

The numerical evaluation of the non-equilibrium CPT proceeds in two steps:
(i) The contour-ordered Green's function $\ff G'$ of the reference system of disconnected clusters has to be calculated. 
If the individual cluster is sufficiently small, this can be done by full diagonalization of $B$ and $H(t)$.
The computation is straightforward. 
(ii) The CPT equation (\ref{eq:dysoncpt2}) must be solved. 
This is a Fredholm integral equation of the second kind which has the same formal structure as Dyson's equation (\ref{eq:dysonlong}).
The standard approach consists in a discretization of the time variables to cast the CPT equation into a matrix form and to employ standard techniques for the solution of inhomogeneous linear systems of equations for its solution. 
It is recommendable to consider the CPT equation (\ref{eq:dysoncpt2}) in the form
\begin{equation} 
  (\ff 1 - \ff G' \cdot \ff V_{\rm K} \otimes \ff 1) \ff G_{\rm CPT} = \ff G'  \: ,
\end{equation} 
as its solution formally requires a single inversion of a well-conditioned matrix only.

We use $N_{\rm M}$ time slices for the Matsubara branch and $N_{\rm K}$ time slices for the upper as well as for the lower branch. 
This leads to a matrix dimension of $N_{\rm M} + 2 N_{\rm K}$.
Using \refeq{cptmat} to separate the solution of the CPT equation on the Matsubara branch from the rest of the problem, leads to $N_{\rm M} \times N_{\rm M}$ and $2 N_{\rm K} \times 2 N_{\rm K}$ matrices only. 
Exploiting further properties of the contour-ordered Green's function, one can reformulate Dyson's equation such that only $N_{\rm M} \times N_{\rm M}$, $N_{\rm K} \times N_{\rm K}$ and $N_{\rm M} \times N_{\rm K}$ matrices must be considered for five independent quantities. \cite{Wag91,MT08}

For the time discretization, a maximal real time $t_{\rm max}$ has to be introduced as a cutoff of the Keldysh contour. 
This can be justified with arguments analogous to those given in the preceding section: 
If $t,t' < t_{\rm max}$, the integrations over $t''$ (and $t'''$) in \refeq{tintegrals} and thus in \refeq{dysonlong} from $t''=t_{\rm max}$ to $t''=\infty$ (upper branch) and from $t''=\infty$ to $t''=t_{\rm max}$ (lower branch) cancel each other. 
Hence, any choice of $t_{\rm max} > t_0$ is justified.
On the other hand, $t_{\rm max}$ determines the maximal observation time up to which the Green's function 
$G_{\rm CPT, \alpha\alpha'}(t,t')$ and thus expectation values $\langle A \rangle_t$ can be calculated.
An immediate consequence of this is that non-equilibrium CPT cannot access the long-time behavior of observables:
The numerical effort is dominated by the solution of linear systems of equations with a dimension $N_{\rm K}$ proportional to $t_{\rm max}$ and therefore increases asymptotically as $t_{\rm max}^3$.
Note, that matrix dimensions also increase due to site and orbital indices.

\subsection{Limiting cases}

Comparing the exact with the CPT self-energy, \refeq{sigma} with \refeq{sigmacpt}, shows that the non-equilibrium CPT becomes exact in the non-interacting limit $\ff U_{\rm K}=0$ as well as in the limit of decoupled clusters $\ff V_{\rm K}=0$.
The latter is, of course, trivial. 
The non-interacting limit, on the other hand, provides a serious check for the numerical evaluation of the theory.

We have performed calculations for the single-band Hubbard model on a linear chain consisting of $L$ sites with open boundaries:
\begin{equation} 
  H = -T \sum_{i=1}^{L-1} \sum_{\sigma=\uparrow,\downarrow} (c_{i\sigma}^\dagger c_{i+1\sigma} + \mbox{H.c.})
  + \frac{U}{2} \sum_{i\sigma} n_{i\sigma} n_{i-\sigma} \: .
\labeq{hub}
\end{equation} 
Here, $T=1$ is the nearest-neighbor hopping which fixes the energy scale.
Using the non-equilibrium CPT for Hubbard interaction $U=0$, we have calculated the site-dependent occupation $\langle n_{i\sigma} \rangle_t = \langle c^\dagger_{i\sigma} c_{i\sigma} \rangle_t = n_i(t)$ as a function of the time $t$ for spin-symmetric conditions.
The initial state, prepared at $t=t_0=0$, is assumed to be a pure state where $N=L$ electrons occupy the sites $i=1,...,L/2$.
This is a half-filled chain with all electrons located on the left half.
Calculations are performed for $L=4$ sites to allow for a check of the CPT results against the exact time evolution of $n_i(t)$.
The reference system is taken to be given by two (non-interacting) Hubbard clusters consisting of two neighboring sites each such that the inter-cluster hopping, which in the CPT is treated perturbatively to all orders, is given by the hopping between the right site of the first and the left site of the second cluster. 

\reffig{check} shows the results for different $\Delta t \equiv t_{\rm max}/N_{\rm K}$. 
Choosing $t_{\rm max}=10$ and $\Delta t = 0.01$ implies $N_{\rm K} = 2000$ time points on the Keldysh branch. 
As can be seen from the figure by comparing with the exact solution, this turns out to be sufficient for convergence of the results.
The figure also demonstrates that the numerical evaluation recovers the $U=0$ limit correctly.
The physics of this example is simple: 
For small $t$, the occupation of the second site quickly decreases, while due to Pauli blocking, the occupancy at the first site starts to decrease with some time delay.
On a larger time scale, a strongly  oscillatory time evolution is observed as it is characteristic for a finite small system.

\begin{figure}[t]
\centerline{\includegraphics[width=0.45\textwidth]{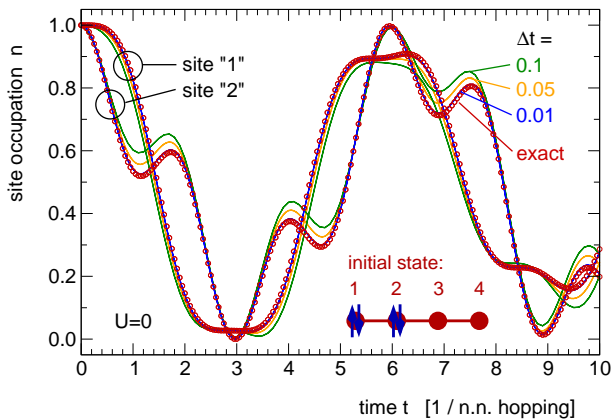}}
\caption{ 
Time dependence of the site occupations in a four-site Hubbard chain for $U=0$.
Results of the non-equilibrium cluster-perturbation theory (lines) are compared with the exact result (points)
for different time discretizations $\Delta t$ on the Keldysh contour.
The initial state is displayed schematically. 
At $t=0$ four electrons occupy the sites 1 and 2 while sites 3 and 4 are empty. 
The CPT treats the nearest-neighbor hopping between sites 2 and 3 perturbatively to all orders.
}
\labfig{check}
\end{figure}

\subsection{Initial-state correlations}

For the above calculations, we only took the Keldysh contour into account and set $\beta=0$. 
This is correct for an initial state represented by a Hamiltonian $\ca B$ with vanishing inter-cluster hopping $\ff V_{\rm B}$ as it is the case here: The initial state is obtained as the ground state of the Hubbard model with vanishing hopping between sites 2 and 3 and suitably chosen on-site energies to realize a filled left and an empty right cluster.

$\ff V_{\rm B}=0$ implies that the Matsubara branch is irrelevant for the time evolution within the CPT.
To prove this, we consider the CPT equation \refeq{dysoncpt2}.
As a matrix in $t,t'$ the Green's function of the reference system consists of four blocks,
\begin{equation}
  \ff G' 
  = 
  \left( 
  \begin{array}{c|c}
  \ff G'_{\rm KK} & \ff G'_{\rm KM} \\
  \hline
  \ff G'_{\rm MK} & \ff G'_{\rm MM}
  \end{array} 
  \right) \; , 
\end{equation}
where K refers to the upper and the lower branches of the Keldysh contour and M to the Matsubara branch.
The block structure for non-retarded, instantaneous potential scattering is simple:
\begin{equation}
  \ff V_{\rm K} \otimes \ff 1
  = 
  \left( 
  \begin{array}{c|c}
  \ff V \otimes \ff 1 & 0 \\
  \hline
  0 & 0
  \end{array} 
  \right) \; .
\end{equation}
The matrix is diagonal and the MM block is zero for an initial state with $\ff V_{\rm B}=0$.
This immediately implies that the KK block of the CPT Green's function satisfies a simplified CPT equation,
\begin{equation}
  \ff G_{\rm CPT, KK} = \ff G'_{\rm KK} + \ff G'_{\rm KK} \cdot \ff V \otimes \ff 1 \cdot \ff G_{\rm CPT, KK} \; ,
\labeq{cptsimple}
\end{equation}
and depends on the KK block of the reference system's Green's function only.

Within general non-equilibrium perturbation theory, the Matsubara branch cannot be disregarded unless the initial state is uncorrelated: \cite{Wag91,MT08}
Only if $\ff B_1 = 0$ there are no vertices with imaginary time entries in the diagrammatic expansion of $\ff G$.
In the presence of initial-state correlations, however, the Matsubara branch is needed to expand the many-body density operator in terms of the non-interacting density operator which is a necessary prerequisite for the application of Wick's theorem.

Within non-equilibrium CPT, on the other hand, interaction vertices generated by $\ff U_{\rm K}$ (including $\ff U_{\rm B}$) are taken into account to all orders for $\ff V_{\rm K}=0$ by the numerically exact calculation of the Green's function of the reference system $\ff G'$.
The subsequent summation of diagrams generated by $\ff V_{\rm K}$, however, can be restricted to vertices on the Keldysh contour only since $\ff V_{\rm B}=0$ is assumed. 
The absence of effects of initial-state correlations on the real time evolution must therefore be seen as an artifact of the CPT.
In fact, the self-energy diagram (c) in \reffig{dyson} is just a prime example to see this:  
We assume the interaction vertices in this diagram to have imaginary time entries, i.e.\ we assume the interaction lines to be labeled by $\ff U_{\rm B}$, which may occur in case of a correlated initial state.
Now, while the diagram is neglected within CPT, it gives a non-vanishing contribution within full perturbation theory even if $\ff V_{\rm B}=0$ since an interaction vertex at imaginary time and a potential-scattering vertex at real time can be connected by a non-vanishing element of the MK block of the free propagator $\ff G'_0$.

\begin{figure}[t]
\centerline{\includegraphics[width=0.45\textwidth]{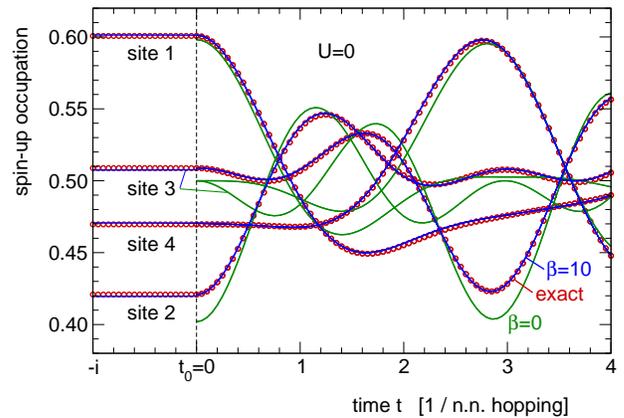}}
\caption{ 
Time dependence of the spin-$\uparrow$ site occupations in a four-site Hubbard chain for $U=0$ as obtained by non-equilibrium cluster-perturbation theory starting from a reference system with decoupled two-site clusters and treating the nearest-neighbor hopping between sites 2 and 3 perturbatively to all orders. 
The initial state is the ground state of the $U=0$ chain at half-filling with a local magnetic field of strength $-h$ at site 1 and $+h$ at site 2 with $h=0.2$.
CPT results for large $\beta = 10$ and for $\beta = 0$ (i.e.\ neglecting the effects of the Matsubara branch).
Exact results (points) are shown for comparison.
}
\labfig{isc}
\end{figure}

\reffig{isc} gives an example for a case where, within CPT, the effect of the Matsubara branch is essential. 
We again consider the $U=0$ Hubbard chain with $L=4$ sites at half-filling. 
The system is assumed to be initially in the ground state of the same model but with an external magnetic field. 
This might also be seen as a magnetic-field quench.
To induce a spatially asymmetric situation, we consider an additional field term
\begin{equation}
  H_{\rm field} = h \sum_{i=1}^2 (-1)^i (n_{i\uparrow} - n_{i\downarrow})
\labeq{field}
\end{equation}
to the Hamiltonian \refeq{hub} which is staggered and non-zero on sites 1 and 2 only.
In the figure, the resulting exact time dependence for $t>t_0=0$ is shown for $\langle n_{i\uparrow} \rangle_t$ as points. 
For the $\sigma=\downarrow$ channel we have $\langle n_{i\downarrow} \rangle_t = 1 - \langle n_{i\uparrow} \rangle_t$.
For $t=0$, the magnetic moments at sites 1 and 2 are considerably larger than those at sites 3 and 4, due to the locally applied field.
For $t=3$, the situation is reversed, and the moments on sites 3 and 4 are larger. 

Since $U=0$ the CPT is expected to provide the exact result. 
In fact, for a reference system with decoupled two-site clusters (1 and 2 decoupled from 3 and 4) the calculation for $\beta = 10$ is close to the exact solution.
This holds for the initial state, as can be seen be comparing the site occupations with the exact ones for $t=0$, as well as for the subsequent time evolution. 
Residual deviations result from the finite time grid with $\Delta t = 0.005$ on the Keldysh and on the Matsubara branch.

In addition, the result of a CPT calculation with $\beta=0$ is shown in \reffig{isc}. 
This corresponds to a calculation on the Keldysh branch only but starting with the same Green's function of the reference system.
Obviously, there are strong deviations from the exact result which proves the relevance of the Matsubara branch for the CPT calculations.
For the present example the initial state is given as the ground state of a Hamiltonian $\ca B$ with $\rm V_{\rm B} = \ff V \ne 0$.
Consequently, the Matsubara branch is required to restore the effect of the inter-cluster potential in the initial state.
Note that a finite field strength is necessary here ($h=0.2$). 
For $h\to \infty$ the two clusters of the initial-state Hamiltonian $\ca B$ would decouple dynamically, and the initial state could be described with $\rm V_{\rm B} = 0$ equivalently, and the Matsubara branch would become irrelevant. 
Furthermore, we note that the $\beta=0$ results correspond to a calculation with $\rm V_{\rm B} = 0$ in the initial-state Hamiltonian $\ca B$ since, as argued above, in that case the simplified CPT equation (\ref{eq:cptsimple}) on the Keldysh contour holds.

\subsection{Exploiting symmetries}

If the time evolution of a pure state $\rho = |{\Psi} \rangle \langle {\Psi}|$ is considered, the symmetries of the contour-ordered Green's function must be taken into account carefully. 
In the CPT calculation, the pure initial state $|\Psi\rangle$ is obtained as the ground state of a suitably chosen Hamiltonian $\ca B$ by exact diagonalization.
$|\Psi\rangle$ is then used to get the reference system (cluster) Green's function $\ff G'$ as an expectation value. 
On the other hand, the Matsubara branch has to be cut off at a finite parameter $\beta<\infty$. 
This implies that for $t \to -i\beta$ the cluster Green's function, obtained as a ground-state expectation value, cannot respect the symmetry relations
\begin{eqnarray}
  G_{\alpha\alpha'}(t,t')  & = & G^*_{\alpha'\alpha}(t',-i\beta - t) \; , \labeq{sym1}\\
  G_{\alpha\alpha'}(t',t)  & = & G^*_{\alpha'\alpha}(-i\beta - t,t') \; , \labeq{sym2}
\end{eqnarray}
which hold exactly for $t$ on the Matsubara and $t'$ on the Keldysh branch and for a Green's function corresponding to a mixed thermal initial state with inverse temperature $\beta<\infty$.
Hence, the limits $\beta \to \infty$ and $t \to -i \beta$ do not commute.
Even for large but finite $\beta$ in the CPT calculation, however, the behavior of $G'_{\alpha\alpha'}(t,t')$ for $t \to -i\beta$ or $t' \to -i\beta$ cannot be neglected, provided that the Matsubara branch is necessary at all, of course. 
The reason is that $G'_{\alpha\alpha'}(t,t')$, considered as a matrix in $t$ and $t'$, does {\em not} adopt a block-diagonal structure in the $\beta\to\infty$ limit.
We therefore enforce the symmetries \refeq{sym1} and \refeq{sym2} by hand:
The expectation value with $|\Psi\rangle$ is calculated for $t=-i\tau$ and $t'=-i\tau'$ with $0<\tau,\tau'<\beta/2$ and the symmetry relations are then used to get $G_{\alpha\alpha'}(t ,t')$ for $\beta/2 < \tau,\tau' <\beta$.
Clearly, for finite $\beta$ this introduces artificial discontinuities of the Green's function at $t = -i \beta/2$ and $t' = -i \beta/2$. 
The height of the jumps, however, disappear asymptotically for $\beta\to\infty$. 
Consequently, it is easily verified numerically that convergence to the exact result can be achieved for large $\beta$ if the symmetries \refeq{sym1} and \refeq{sym2} are enforced while strong deviations from the exact result remain present even for $\beta\to\infty$ otherwise. 

For efficiency reasons, one may exploit more symmetry relations. 
In fact, we find it convenient to profit from the exact relations
\begin{eqnarray}
  G_{\alpha\alpha'}(t_\lo,t'_\up) & = & -G^*_{\alpha'\alpha}(t'_\lo,t_\up) \; , \labeq{sym3}\\
  G_{\alpha\alpha'}(t_\up,t'_\lo) & = & -G^*_{\alpha'\alpha}(t'_\up,t_\lo) \; , \labeq{sym4}
\end{eqnarray}
which hold for $t,t'$ on the Keldysh branch. 
Here, $t_\up$ indicates that $t$ belongs to the upper branch while $t_\lo = t$ but lies on the lower branch. 
We also make use of time homogeneity on the Matsubara branch, 
\begin{equation}
  G_{\alpha\alpha'}(t,t') = G_{\alpha\alpha'}(t-t') \: ,
\end{equation}
valid for $t=-i\tau$, $t'=-i\tau'$ and $0<\tau,\tau'<\beta$.

\subsection{Short-time dynamics}

\reffig{short} shows the results of a calculation for the Hubbard model \refeq{hub} in the strong-coupling regime for $U=8$.
To allow for a comparison of the results from non-equilibrium CPT with the exact results, we consider the $L=4$ site chain again.
Initially, the system is prepared in the N\'e{}el state $| \Psi \rangle$ where $\langle \Psi | n_{i\uparrow} | \Psi \rangle = 1$ for $i=1$ and $i=3$ and $\langle \Psi | n_{i\uparrow} | \Psi \rangle = 0$ for $i=2$ and $i=4$ and where
$\langle \Psi | n_{i\downarrow} | \Psi \rangle = 1 - \langle \Psi | n_{i\uparrow} | \Psi \rangle$.
For strong $U$ at half-filling the Hubbard model maps onto the antiferromagnetic Heisenberg model with a ground state and excited energy eigenstates that are different from the classical N\'e{}el state. 
This induces a non-trivial dynamics as can be seen from the exact calculation (blue lines) in \reffig{short}.

The N\'e{}el state may be obtained as the ground state of an initial-state Hamiltonian $\ca B$ with a staggered magnetic field term as in \refeq{field} but applied to all sites and with field strength $h\to \infty$. 
This implies that the sites are decoupled dynamically, and that $\ff V_{\rm B}=0$ can be assumed for the initial state. 
Consequently, we are allowed to disregard the Matsubara branch.

For the CPT calculation we again start from a reference system with decoupled two-site clusters. 
By construction, the initial state is described correctly within the CPT approach. 
As can be seen from \reffig{short}, the site occupations obtained by CPT (red lines) deviate from the exact results for $t>0$ as expected for $U\ne 0$.
For comparison, the time dependence of the site occupations of the reference system are given in addition (green line). 
The reference system has a higher symmetry which leads to occupations of sites 1 and 2 that are related to each other by spin reversal.
For larger times the CPT results seem to follow more or less the time dependence of the site occupations in the reference system. 
This means that the approximation is not able to describe the effects of inter-cluster correlations correctly and that intra-cluster effects dominate the behavior of $\langle n_{i\sigma} \rangle_t$ at large times.

On the other hand, at short times $t \lesssim 0.3$, the CPT results are clearly different from the site occupations of the reference system and to a high precision follow the exact trend.
We conclude that inter-cluster correlations, as represented by the diagram (c) in \reffig{dyson}, are ineffective at short times even if the interaction is strong. 
The fact that the non-equilibrium CPT describes the short-time dynamics of single-particle operators exactly, is interpreted to be the analog of the fact that the equilibrium CPT predicts global, i.e.\ frequency-integrated properties of the single-particle excitation spectrum correctly. 
The CPT apparently respects to a good approximation the first few non-equilibrium moment sum rules which determine the short-time dynamics. \cite{TF06,TF08} 

\begin{figure}[t]
\centerline{\includegraphics[width=0.45\textwidth]{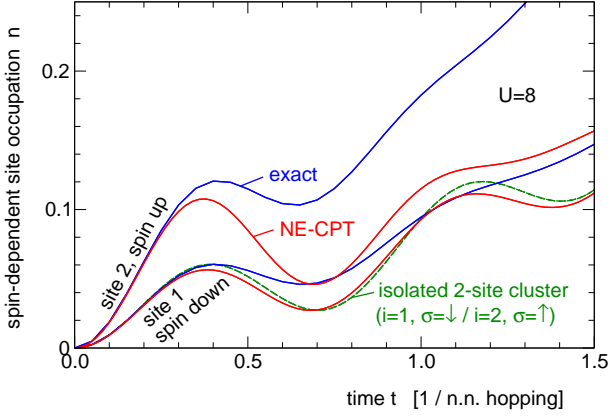}}
\caption{
Time evolution as obtained from non-equilibrium CPT (red), 
exact time evolution (blue) 
and time evolution of the reference system (green)
in the strong-coupling regime ($U=8$) of a half-filled four-site Hubbard chain with a N\'eel initial state.
Reference system: decoupled two-site clusters. 
Inter-cluster potential: nearest-neighbor hopping connecting sites 2 and 3. 
}
\labfig{short}
\end{figure}

\subsection{Coupling to an infinite bath}

The (equilibrium) CPT has actually been designed to treat correlated electrons on an infinite lattice.
For the non-equilibrium case, the results presented above represent simple test calculations which demonstrate that two correlated clusters with $L_c$ sites each can be coupled to a single but larger cluster with $2L_c$ sites.
This scheme can be iterated straightforwardly to build up extended lattices with or without translational symmetries.

Besides this, the non-equilibrium CPT can also be used to couple a small correlated ``system'' to an uncorrelated ``bath'' with a large number of degrees of freedom, such as a magnetic nanostructure on a metal surface or a molecule coupled to external leads etc.
We assume that the Hilbert-space dimension of the correlated system is sufficiently small such that the contour-ordered Green's function $\ff G'_{\rm S}$ can be calculated exactly.
By means of \refeq{g01} and \refeq{g02} we also have the Green's function of the bath $\ff G'_{\rm B}$ for an in principle arbitrarily large number of uncorrelated bath sites.
Hence, the Green's function of the decoupled reference system, given by $B'$ and $H'(t)$, can be written as a $2\times 2$ matrix
\begin{equation}
  \ff G' 
  = 
  \left( 
  \begin{array}{c|c}
  \ff G'_{\rm S} & 0 \\
  \hline
  0 & \ff G'_{\rm B}
  \end{array} 
  \right) \; ,
\end{equation}
with entries referring to system or bath orbitals. 
The coupling of the system to the bath is provided by the ``inter-cluster'' term, i.e.\ by the hybridization
\begin{equation}
  \ff V_{\rm K} \otimes \ff 1
  = 
  \left( 
  \begin{array}{c|c}
  0 & \ff V_{\rm K} \\
  \hline
  \ff V^\dagger_{\rm K} & 0
  \end{array} 
  \right) \otimes \ff 1
  \; .
\end{equation}
Within non-equilibrium CPT, the Green's function of the entire system,
\begin{equation}
  \ff G_{\rm CPT}
  = 
  \left( 
  \begin{array}{c|c}
  \ff G_{\rm CPT, S} & \ff G_{\rm CPT, SB} \\
  \hline
  \ff G_{\rm CPT, BS} & \ff G_{\rm CPT, B}
  \end{array} 
  \right) \; ,
\end{equation}
is obtained from the general CPT equation, $\ff G_{\rm CPT} = \ff G' + \ff G' \cdot \ff V_{\rm K} \otimes \ff 1 \cdot \ff G_{\rm CPT}$ (see \refeq{dysoncpt2}).
With the definition of the non-equilibrium hybridization function,  
$\ff \Delta = \ff V_{\rm K} \otimes \ff 1 \cdot \ff G'_{\rm B} \cdot (\ff V_{\rm K} \otimes \ff 1)^\dagger$,
or, in short,
\begin{equation}
  \ff \Delta
  = 
  \ff V \cdot \ff G'_{\rm B} \cdot \ff V^\dagger
  \; ,
\end{equation}
the CPT Green's function of the system's degrees of freedom is obtained as:
\begin{equation}
  \ff G_{\rm CPT, S}
  = 
  \frac{\ff 1}{{\ff G_{\rm S}'}^{-1} - \ff \Delta}
  \; .
\labeq{impcpt1}
\end{equation}
For the situation considered here, this can be seen as a simplified CPT equation which is decoupled from the remaining CPT equations for the bath and system/bath Green's functions,
\begin{equation}
  \ff V \cdot \ff G_{\rm CPT, B} \cdot \ff V^\dagger 
  =  
  \frac{\ff 1}{\ff \Delta^{-1} - \ff G_{\rm S}'}
\labeq{impcpt2}
\end{equation}
and
\begin{eqnarray}
  \ff V \cdot \ff G_{\rm CPT, BS} 
  &=& 
  \ff \Delta \cdot \ff G_{\rm CPT, S} 
  \; , \nonumber \\
  \ff G_{\rm CPT, SB} \cdot \ff V^\dagger 
  & = &
  \ff G'_{\rm S} \cdot ( \ff V \cdot \ff G_{\rm CPT, B} \cdot \ff V^\dagger )
  \; .
\labeq{impcpt3}
\end{eqnarray}
Eqs.\ (\ref{eq:impcpt1}) -- (\ref{eq:impcpt3})
have simple diagrammatic representations known from scattering theory. 

For the following a small system with $L_c$ sites is considered and a hybridization that links the site $i_0$ of the system to a single site $i_{0,\rm bath}$ of the bath. 
Let $\rho_0(\varepsilon)$ be the non-interacting local density of states of the bath at $i_{0,\rm bath}$.
This implies that the hybridization function $\ff \Delta$ is non-zero at $i_0$ only, and $\Delta(t,t') \equiv \Delta_{i_0i_0}(t,t') = V(t) G'_{\rm B, loc.}(t,t') V(t')$ where the local bath Green's function at $i_{0,\rm bath}$ is given by
\begin{equation}
  i G'_{\rm B, loc.}(t,t')
  = 
  \int d\varepsilon \, \rho_0(\varepsilon) 
  \frac{ e^{-i(\varepsilon-\mu)(t-t')} }{1+e^{-\beta(\varepsilon - \mu)}}
\labeq{ba1}
\end{equation}
if $t$ is ``later'' then $t'$ on the contour, and
\begin{equation}
  i G'_{\rm B, loc.}(t,t')
  = -
  \int d\varepsilon \, \rho_0(\varepsilon) 
  \frac{ e^{-i(\varepsilon-\mu)(t-t')} }{e^{\beta(\varepsilon-\mu)}  + 1}
\labeq{ba2}
\end{equation}
if $t'$ is ``later'' then $t$ on the contour. 
This means that the bath is fully characterized by its local density of states $\rho_0(\varepsilon)$ at $i_{0,\rm bath}$.
The CPT equation \refeq{impcpt1} then provides the system's Green's function at $i_0$:
\begin{equation}
  \ff G_{\rm CPT, S, i_0i_0} 
  = 
  \frac{\ff 1}{{\ff G'}^{-1}_{\rm S, i_0 i_0} - \ff \Delta} \: ,
\end{equation}
where fat quantities are matrices in $t,t'$ only.
For the other sites we have:
\begin{equation}
  \ff G_{\rm CPT, S, ij} 
  = 
  \ff G'_{\rm S, ij}
  + 
  \ff G'_{\rm S, ii_0}
  \ff \Delta
  \frac{\ff 1}{\ff 1 - \ff G'_{\rm S, i_0 i_0} \ff \Delta}
  \ff G'_{\rm S, i_0j} \; .
\end{equation}

For the numerical calculations we consider a system in a linear geometry with $L_c \ge 2$ sites.
The Hubbard interaction $U$ is non-zero at sites $i=1$ and $i=2$ only, and the hopping between nearest neighbors is set to $T=1$ to fix the energy and time scales.
System sizes range from $L_c=2$ to $L_c=8$. 
The latter is the maximum size that can conveniently be treated by means of exact diagonalization.
Via non-equilibrium CPT this system is coupled at the site $i=i_0=L_c$ to a bath with a semi-elliptic density of states $\rho_0(\varepsilon)$ of bandwidth $W=4T$. 
This is just the local density of states at the first site $i_{0,\rm bath}$ for a semi-infinite linear chain.
Both, the system and the bath, are taken to be at half-filling, i.e.\ we set the chemical potential $\mu=0$, and assume vanishing on-site energies for all sites except for $i=1,2$ where the on-site energy is $-U/2$.
In the ground state $n_\sigma \equiv \langle n_{i\sigma} \rangle = 0.5$ for system and bath.

However, the initial state is taken to be the ground state of another Hamiltonian $B$ which differs from $H$ by (i) the hopping between sites $i=2$ and $i=3$. This hopping is suddenly switched on at time $t=0$. Furthermore, (ii) the correlated two-site model for the initial state is perturbed by either a spin or by a charge excitation. 
This is realized by applying a respective staggered field term:
\begin{equation}
B = B(0) 
- h_{\rm spin} (n_{1\uparrow} - n_{1\downarrow})
+ h_{\rm spin} (n_{2\uparrow} - n_{2\downarrow})
\labeq{spinexc}
\end{equation}
or 
\begin{equation}
B = B(0) 
- h_{\rm charge} (n_{1\uparrow} + n_{1\downarrow})
+ h_{\rm charge} (n_{2\uparrow} + n_{2\downarrow}) \; .
\labeq{chargeexc}
\end{equation}
The Hamiltonians of the initial ground state and of the transient final state, i.e.\ $B$ and $H$, are shown schematically in \reffig{system}.
Note that the CPT describes the initial state exactly because the correlated sites are decoupled and because the coupling of the rest of the sites of the system to the bath is taken into account exactly via CPT since these sites are uncorrelated.
Converged results are obtained with the choice $\beta=5$ for the Matsubara branch. 

\begin{figure}[t]
\centerline{\includegraphics[width=0.35\textwidth]{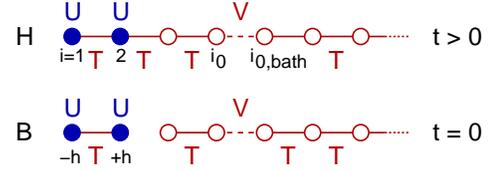}}
\caption{
Pictorial representation of the final-state Hamiltonian $H$ governing the time evolution and the Hamiltonian $B$ generating the initial state as its ground state. 
The {\em system} size is $i_0=L_c$.
The {\em bath} is a semi-infinite linear chain starting at $i_{0,\rm bath}=L_c+1$.
The coupling between system and bath $V$ is treated within CPT ($V=T$).
Blue circles represent sites with $U>0$ while open red circles stand for sites with $U=0$. 
Initially ($t=0$) there is a ground state of a an isolated two-site Hubbard cluster perturbed by a staggered field $h$ coupling to spin or charge degrees of freedom. 
The field is suddenly switched off, and the coupling of the two-site cluster to the rest of the system is switched on. 
Due to particle-hole symmetry, the entire system (``system'' plus bath) is and stays at half-filling.
}
\labfig{system}
\end{figure}

\begin{figure}[t]
\centerline{\includegraphics[width=0.45\textwidth]{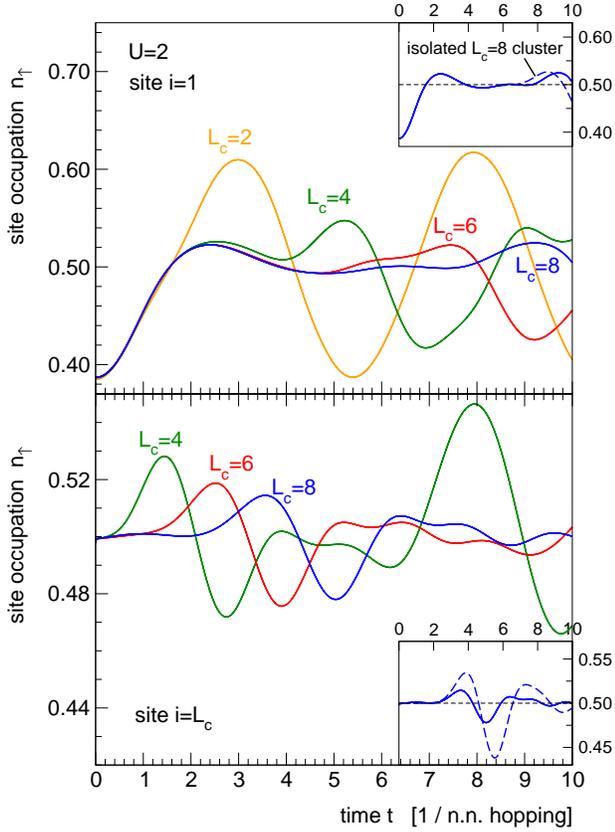}}
\caption{
Time evolution of the average occupation $n_\uparrow = \langle n_{i\uparrow} \rangle = 1 - \langle n_{i\downarrow} \rangle$ at site $i=1$ (upper) and site $i=L_c$ (lower panel) of the system displayed in \reffig{system} after a spin excitation using a staggered field $h_{\rm spin}=0.1$ in the initial-state Hamiltonian (see \refeq{spinexc}). 
Insets: Comparison of the CPT result for $L_c=8$ with the corresponding result obtained from a calculation for an isolated system without bath. The CPT calculations have been performed for different system size $L_c$ as indicated. 
Further parameters: $U=2$, half-filling, $T=V=1$.
}
\labfig{spin}
\end{figure}

\begin{figure}[t]
\centerline{\includegraphics[width=0.45\textwidth]{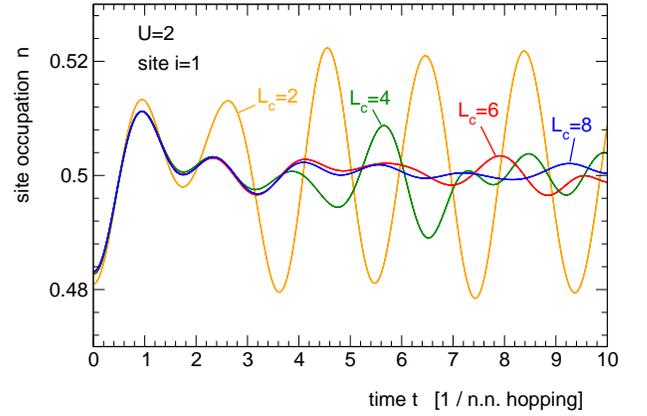}}
\caption{
Time evolution of the average occupation $n = \langle n_{i\uparrow} \rangle = \langle n_{i\downarrow} \rangle$ at site $i=1$ of the system displayed in \reffig{system} after a charge excitation with $h_{\rm charge}=0.1$ (see \refeq{spinexc}). 
System size $L_c$ as indicated, $U=2$, half-filling, $T=V=1$.
}
\labfig{charge}
\end{figure}

Physically, we expect that the initial local perturbation propagates through the chain and dissipates into the bath such that the system relaxes to its ground state with $n_\sigma=0.5$. 
For $U=2$ \reffig{spin} displays the result of a calculation for a spin excitation with $h_{\rm spin}=0.1$, and \reffig{charge} the results for a charge excitation with $h_{\rm charge}=0.1$.
We find that the results for the time dependence of $n_\uparrow(t) = 1 - n_{\downarrow}(t)$ improve with increasing size of the system. 
Clearly, if the CPT was exact there should not be any differences between the results of calculations for different $L_c$. 

At site $i=1$ (\reffig{spin}, upper panel) the result obtained for the smallest system with $L_c=2$ shows a strongly oscillating trend with hardly any damping despite the presence of the bath. 
Here, the non-equilibrium CPT appears reliable on a short time scale $t \lesssim 2$ only as can be seen by comparing with $L_c \ge 4$.
By comparing with the result for the largest size, it is obvious that this time scale rapidly grows with increasing system size.
For $L_c=8$ and up to the accessible maximal time $t_{\rm max}=10$, the trend of $n_\uparrow(t)$ follows our expectation:
The initial spin polarization quickly decreases and, apart from weak remaining oscillations, approaches $n_\uparrow - n_\downarrow \approx 0$.

The lower panel of \reffig{spin} shows $n_\uparrow(t)$ at site $i=i_0=L_c$. 
As the local spin excitation requires a finite time to propagate to $i=i_0$, the results for the different system sizes show a response that is more and more delayed with increasing $L_c$.
The excitation is nicely seen to propagate through $i_0$ and leaving the site in an almost unperturbed state thereafter.
The upturn of $n_\uparrow(t)$ for $t \gtrsim 7$ in the calculation for $L_c=4$, however, must be interpreted as an artifact. 
Here the system size is still insufficient to predict the correct trend up to $t_{\rm max}$.
On the other hand, one should note that the decrease in the amplitude of the response with increasing $L_c$, and thus with increasing distance from the initial perturbation, is reasonable.

\reffig{charge} presents the time-evolution of $n(t)=n_\uparrow(t)=n_\downarrow(t)$ after an initial charge excitation at $i=1$. 
Again, the initially strong deviation from the equilibrium value $n(t) = 0.5$ is quickly dissipated to the bath while the remaining low-amplitude oscillations are expected to decay on a time scale beyond $t_{\rm max}$. 
We also find that the CPT results rapidly improve with increasing system size $L_c$.

Comparing the results for spin and charge excitations, we note that the system is substantially more susceptible to a staggered field that couples to the spin as compared to a field coupling to the charge degrees of freedom;
for $h_{\rm charge} = h_{\rm spin}$ we find oscillations with larger amplitudes in \reffig{spin}.
Furthermore, the characteristic frequency of the oscillations seen in Fig.\ \ref{fig:spin} for the spin excitation is clearly smaller than the corresponding one for the charge excitation (Fig.\ \ref{fig:charge}).
These facts are strongly dependent on $U$.
With increasing $U$ we find that the characteristic frequency for the spin excitation is roughly given by $\omega_{\rm spin} \sim 1/U$ which corresponds to the low-energy Heisenberg scale, while for the charge excitation $\omega_{\rm charge} \sim U$ which corresponds to the high-energy Hubbard bands.
This is accompanied by an increase (decrease) of the amplitudes for the oscillations following the spin (charge) excitation.
For strong $U$, the system is very weakly susceptible to a perturbation coupling to the charge as compared to the spin degrees of freedom.

We conclude that the non-equilibrium CPT is in fact able to describe the dynamics following a perturbation of a small correlated system in a non-interacting environment and the dissipation of a local spin or charge excitation to a large uncorrelated bath. 
It is important to note, however, that the above-mentioned effects are to some extent already captured in a calculation for $V=0$ in the final state, i.e.\ in a calculation without bath. 
This is most apparently seen in the inset of the upper panel in \reffig{spin}, where $n_\uparrow(t)$ obtained by CPT is compared with the result for the isolated cluster at $L_c=8$. 
The CPT does improve the calculation for the isolated cluster but the gain is small. 
The reason is that the ``reflection'' of the propagating excitation at the boundary $i=i_0$ and the back-propagation to $i=1$ takes a time close to $t=t_{\rm max}$.
On the other hand, at site $i=i_0$ (see lower panel), the CPT substantially improves the isolated-cluster calculation by predicting a much stronger damping. 

These observations can also be understood from the diagrammatic construction of the non-equilibrium CPT (see \reffig{dyson}) by assuming that non-diagonal elements of the free intra-cluster propagator, $G'_{0,ij}$ with $i\ne j$ but $i,j\in I$ are small compared to diagonal elements $G'_{0,ii}$ and decrease with increasing distance $|i-j|$. 
The diagram to the self-energy in \reffig{dyson}c, neglected within the CPT, necessarily involves two non-diagonal propagators $G'_{0,ij}$ with $i=1$ or $i=2$ and $j=i_0$ since the $U$ and the $V$ vertex are local and separated by a distance $i_0-1$ (see \reffig{system}).
It is therefore of the order $\ca O ({G'}^2_{0,ii_0})$ and vanishes with $i_0\to \infty$.
The same argument can be given for any vertex-correction diagram and hence the CPT becomes exact in the limit of $L_c\to \infty$, as expected. 

Likewise, we can argue that the contribution of neglected vertex-correction diagrams to the site occupation at $i=1$ or $i=2$ are of the order $\ca O ({G'}^2_{0,ii_0})$. 
On the other hand, for $i=i_0$, the CPT provides a more reliable estimate since vertex corrections are already of the order $\ca O ({G'}^4_{0,ii_0})$ because of the necessary two additional non-diagonal propagators. 

\section{Conclusions and outlook}
\label{sec:con}

Usually, the cluster-perturbation theory is seen as the first non-trivial level in a systematic strong-coupling expansion, i.e.\ an expansion in the inter-cluster hopping $\ff V$ around a state with decoupled clusters but finite and arbitrarily strong Hubbard-type interaction $\ff U$.
Here, we have shown that the same CPT can be recovered strictly within the framework of weak-coupling perturbation theory. 
This is achieved by formally summing up certain classes of diagrams that are generated when treating $\ff V$ {\em and} $\ff U$ perturbatively.
In this way the CPT can be interpreted as an approximation that neglects vertex corrections, i.e.\ the influence of scattering at the inter-cluster potential on the renormalization of propagators due to the interaction.
One of the benefits of this reformulation is that therewith one can straightforwardly extend the CPT to study the real-time dynamics of systems far from equilibrium.
One simply has to copy the formalism and paste it to the Keldysh-Matsubara time contour.
This defines the non-equilibrium CPT studied here.

The non-equilibrium CPT is characterized as follows:
(i) It comprises the conventional CPT for the description of the initial thermal state and fully reduces to conventional CPT in the case of thermal equilibrium, i.e.\ for the case where the Hamiltonian $H(t)$ that determines the time evolution is assumed to be time independent and set equal to the Hamiltonian $B$ that defines the initial thermal state.

(ii) The non-equilibrium CPT respects the physical consequences of the causality principle: 
Within the CPT the time evolution of the system depends on the initial state preparation but not vice versa.

(iii) The approach is rather flexible and can be applied to a large class of models, namely lattice fermion (or boson) models with local Hubbard-type interactions including impurity models such as the single-impurity Anderson model. 
For bosons, however, the treatment of the condensate phase requires additional efforts. \cite{KAvdL11}
Furthermore, systems with non-local interactions, like a nearest-neighbor density interaction cannot be treated without further approximations, such as a mean-field decoupling of inter-cluster interaction terms. \cite{AEvdLP+04}

(iv) Due to the necessity to solve a generalized CPT equation for time-inhomogeneous Green's functions, operations involving objects indexed with two discretized time variables have to be performed. 
This limits the numerical evaluation of the scheme to short and intermediate time scales in practice.
On the other hand, there are in principle no limitations concerning the time dependence of the Hamiltonian, and the non-equilibrium CPT can likewise treat sudden parameter changes or periodically driven systems, for example.

(v) The neglect of vertex corrections represents a severe approximation.
This approximation is in principle controlled, however, by the cluster size, i.e.\ the (non-equilibrium) CPT approximation improves with increasing $L_c$. 
This is shared with the conventional (thermal) CPT and classifies the scheme as a cluster mean-field approach where correlations are treated exactly up to the cluster extension and treated in a mean-field way beyond this scale. 
For impurity-type models with a single or a few correlated sites and a continuum of uncorrelated bath degrees of freedom, the approximation has also been seen to improve with increasing distance of the correlated sites from the cluster boundary. 
Here, ``improvement'' means that the dynamics of expectation values of single-particle observables can be traced reliably on longer and longer time scales.
On a very short time scale, the non-equilibrium CPT has been found to recover the exact solution, i.e.\ it apparently (like the equilibrium CPT) respects the first non-trivial moment sum rules. 

(vi) The non-equilibrium CPT can also be characterized as a scheme that interpolates between the isolated-cluster limit ($\ff V=0$) and the band limit ($\ff U = 0$) which are recovered exactly. 
However, already at the second order in the interaction strength there are diagrams missing. 
An interesting case that should be accessible to the method and has been studied experimentally, for bosonic atoms in optical lattices, \cite{ALB+07,FTC+07} are weakly coupled double wells or weakly coupled plaquettes.

Concluding, the approach represents a very flexible and easy to handle method with very moderate computational cost that can give a first access to a rather broad class of systems of strongly correlated electrons far from equilibrium. 
On the other hand, its main drawbacks consist in the missing self-consistency, the neglect of correlations beyond the cluster size and also the artificial breaking of lattice symmetries.
The present work has presented a number of test calculations. 
These can be improved in various ways to achieve more reliable results: 
Larger clusters can be taken into account by replacing the exact-diagonalization approach for the computation of the Keldysh Green's function with e.g.\ a time-adaptive Krylov construction. \cite{PL86}
A (strong-coupling) diagrammatic expansion around the non-equilibrium CPT may be used \cite{JLB+10} to include some of the neglected vertex corrections. 
Alternatively, one can also attempt to enlarge the class of diagrams considered in the presented weak-coupling expansion.
Finally, an optimization of intra-cluster one-particle parameters can be envisaged to introduce a self-consistent feedback within the method which is necessary to study phase transitions and to make contact with non-equilibrium dynamical mean-field theory, for example.
This can be accomplished by a suitable generalization of the self-energy-functional approach.
Work along these lines is in progress.  \cite{PB11}

\acknowledgments
We would like to thank A. I. Lichtenstein for instructive discussions.
The work is supported by the Deutsche Forschungsgemeinschaft within the Sonderforschungsbereich 668 (projects A14 and B3) and by the Cluster of Excellence Nanospintronics (LExI Hamburg) and has been inspired by the SFB925 initiative.

\bibliography{../../lit/lit.bib}

\end{document}